\documentclass{WileyMSP-template}
\begin{document}

\pagestyle{fancy}

\title{Radioactive ion beam production at the Gamma Factory}

\maketitle


\author{D.\ Nichita$\,^{1,2}$}
\author{D.L.\ Balabanski$\,^{1,2}$}
\author{P.\ Constantin$^*\,^1$}
\author{M.W.\ Krasny$\,^{3,4}$}
\author{W.\ P{\l}aczek$\,^5$}



\begin{affiliations}
${}^1\,$Extreme Light Infrastructure -- Nuclear Physics, ``Horia Hulubei'' National Institute for R\&D in Physics \\ ~~and Nuclear Engineering, 077125 Bucharest-Magurele, Romania\\
${}^2\,$Doctoral School in Engineering and Applications of Lasers and Accelerators, University Polytechnica  \\
~~of Bucharest, 060042 Bucharest, Romania\\
${}^3\,$LPNHE, Sorbonne Universit\'e, Universit\'e de Paris, CNRS/IN2P3, Tour 33, RdC, 4, pl. Jussieu,\\ ~~75005 Paris, France\\
${}^4\,$CERN, Esplanade des Particules 1, 1211 Geneva 23, Switzerland\\
${}^5\,$Institute of Applied Computer Science, Jagiellonian
University, ul.\ {\L}ojasiewicza 11, 30-348 Krak\'ow, \\
~~Poland \vspace{1mm}\\
${}^*\,$Corresponding author: paul.constantin@eli-np.ro\\
\end{affiliations}



\begin{abstract}

A very intense $\gamma$ beam of the Gamma Factory facility proposed at CERN can be used to generate radioactive ion beams (RIBs) with high production yields and study the structure of exotic neutron-rich nuclei. The radioactive nuclides are generated via photo-fission in several actinide targets and thermalized in high-purity cryogenic helium, filling a gas cell which is enclosing the targets. Electric fields are used to extract heavy ions and form RIBs which can be send to various selection and measurement stations. Estimates for the production and extraction yields of exotic neutron-rich nuclei with such a setup are provided. A study of the impact of space charge, build-up inside the gas cell, on the extraction properties is presented and it is demonstrated that the beam needs to be chopped for achieving optimal extraction yields.

\end{abstract}


\section{Introduction}

The Gamma Factory (GF) facility at CERN \cite{Krasny:2015ffb,GF-PoP-LoI:2019} can provide the next generation $\gamma$ beams, based on resonant absorption and emission of laser photons on partially stripped ultra-relativistic heavy-ion (PSI) beams. The state-of-the art $\gamma$-beam facilities, such as HI$\gamma$S at Duke University, USA~\cite{Weller:2009} or the VEGA system at ELI-NP, which is in a final stage of construction at Magurele, Romania, are based on Compton back-scattering of laser photons off relativistic electrons (LCB). The $\gamma$-beam flux at the GF is expected to be several orders of magnitude higher compared to the present generation $\gamma$-beam facilities. This is mainly due to the interaction cross section which is higher by up to nine orders of magnitude for the absorption of laser photons by PSIs than that for LCB.

One of the potential applications of this unprecedented $\gamma$-beam intensity is the generation of high yield radioactive-ion beams (RIBs) via photon induced fission. The fission process has been used successfully in the production of intermediate mass ($A \sim 70\,$--$\,150$) nuclides in the neutron-rich region far away from the valley of $\beta$ stability. The study of exotic nuclides in this region is the test-bench for theory in areas like the nuclear equation of state, nuclear structure models and nucleosynthesis via the rapid neutron capture process (\emph{r}-process). Recent measurements of $\gamma$-radiation spectra emitted from neutron star mergers \cite{Kasen:2017} have indicated that these cosmic events are one of the likely locations where \emph{r}-process nucleosynthesis takes place. This has increased the interest in studying the neutron-rich nuclides along the \emph{r}-process path, with special interest around the waiting points at neutron numbers $N = 50$, $N = 82$ and $N = 126$, the first two being accessible in fission, while the last is reached in fragmentation or multi-nucleon transfer reactions.

Several RIB facilities are currently active worldwide, such as CARIBU (ANL) \cite{Savard:2008}, ISAC (Triumf) \cite{Sen:2016}, ISOLDE (CERN) \cite{Voulot:2008}, FRS (GSI) \cite{Plass:2013}, SPIRAL (GANIL) \cite{Delahaye:2020}, JYFL (Jyv\"askyl\"a) \cite{Aysto:2001}, and RIBF (RIKEN) \cite{Sumikama:2016}. For a recent review see Reference \cite{Blumenfeld:2013}. To a large degree, they complement each other by employing a variety of methods and technologies, such as beam types (from heavy ions to photons), target types (thick or thin), fragment separation (in-flight separators or in-cell catchers) and selection (with lasers, magnets, time-of-flight spectrometers), and experimental stations. Among the RIB facilities with $\gamma$-driver beams and thick actinide targets, like the one discussed here, the current ALTO (IPN Orsay) \cite{Mhamed:2020} and the future ARIEL (TRIUMF) \cite{Babcock:2020} facilities employ bremsstrahlung sources, while the future ELISOL facility at ELI-NP uses a LCB source. When compared to facilities with charged particle or ion driver beams, the facilities with the $\gamma$-driver beams have the advantage of negligible background effects due to their primary beam, but also the disadvantage of typically lower RIB production yields. As it will be argued in this paper, a GF-based RIB facility has very high production yields, while keeping background effects generated by the primary beam, like the space charge effect, at a very low level.

The work presented here discusses the opportunity to develop a RIB setup using the GF PSI source as a driver beam. Section~2 describes the GF source generation and its properties relevant to photo-fission inside gas cells. Section~3 details the experimental setup proposed for the RIB formation and the selection of neutron-rich nuclides. Section~4 presents simulations used to estimate the production yields and extraction efficiencies for this setup. 
Finally, a summary of this work is provided in Section~5.

\section{Beam generation and properties}

Photon emission in the helium-like calcium PSI beam collisions with the laser pulses at the GF facility has been simulated with the Monte Carlo event generator {\sf GF-CAIN} \cite{GF-CAIN,GF-PoP-LoI:2019} using the input parameters given in Table~\ref{tab:CainPars}. {\sf GF-CAIN} is a version, adapted for the Gamma Factory, of the simulation code {\sf CAIN}~\cite{CAIN} developed at KEK (Japan) for beam--beam interactions at the International Linear Collider (ILC). The interactions between laser photons and PSI beams in {\sf GF-CAIN} include an atomic absorption of laser photons by a helium-like $\mathrm{Ca}$ ion and its subsequent de-excitation by spontaneous or stimulated emissions, including a finite lifetime of the ion in the excited state. In the laboratory reference frame, this lifetime ($\sim 14\,$ps) is shorter than the laser pulse duration ($50$ or $500\,$ps), hence a single $\mathrm{Ca}$ ion can be excited many times during the bunch-crossing, and as a result can emit multiple photons. On average, the number of emitted photons per ion is in the range $5.5\,$--$\,38.6$ for the laser pulse duration between $50$ and $500\,$ps with the pulse energy between $0.5$ and $2.0\,$mJ. To achieve the requisite pulse energies at the 20 MHz repetition rate the
Fabry-P\'erot cavity has to be used. 

\begin{table}[!htbp]\centering
\caption{Input parameters for the {\sf GF-CAIN} simulations of RIBs at the Gamma Factory}
	\begin{tabular}{lr}\hline\hline
		PSI beam  & ${}^{40}_{20}\mathrm{Ca}^{18+}$\\
		\hline
		$m$ -- ion mass & $37.332\,$GeV/c$^2$ \\
		$\gamma_L=E/mc^2$-- mean Lorentz relativistic factor & $2394.9782$ \\
		$E$ -- mean energy & $89.4093$ TeV \\
		$N$ -- number ions per bunch & $3\times 10^9$ \\
		$\sigma_E/E$ -- RMS relative energy spread & $2\times 10^{-4}$  \\
		$\epsilon_x=\epsilon_y$ -- geometric transverse emittance & $3\times 10^{-10}\;$m$\times$rad\\
		$\sigma_x=\sigma_y$ -- RMS transverse size & $0.1225\,$mm\\
		$\sigma_z$ -- RMS bunch length & $15\,$cm\\
		Bunch repetition rate & $20\,$MHz\\
		\hline
		Laser & Er:glass\\
		\hline
		$\lambda$ -- wavelength ($\hbar\omega$ -- photon energy) & $1521.84\,$nm ($0.8147\,$eV) \\
		$\sigma_{\lambda}/\lambda$ -- RMS relative band spread & $2\times 10^{-4}$  \\
		$U$ -- single pulse energy at IP & $0.5\,$mJ ($2\,$mJ)\\
		$\sigma_x=\sigma_y$ -- RMS transverse intensity distribution at IP & $0.15\,$mm\\
		$\sigma_t$ -- RMS pulse duration & $50\,$ps ($500\,$ps) \\
		$\theta_l$ -- collision angle & $0\degree$ \\
		\hline
 		 Atomic transition of ${}^{40}_{20}\mathrm{Ca}^{18+}$ & $1s^2\;{}^1\mathrm{S}_0 \rightarrow 1s\,2p\;{}^1\mathrm{P}_1$\\
		\hline
		$\hbar\omega'_0 $ -- resonance energy & $3902.3775\,$eV\\
		$\tau'$ -- mean lifetime of spontaneous emission & $6\,$fs\\
		$\hbar\omega_{\gamma}^{\rm max}$ -- maximum emitted photon energy & $18.692\,$MeV\\
		\hline\hline
	\end{tabular}	
\label{tab:CainPars}	
\end{table}

\begin{figure}[!htbp]\centering
  \includegraphics[width=0.6\linewidth]{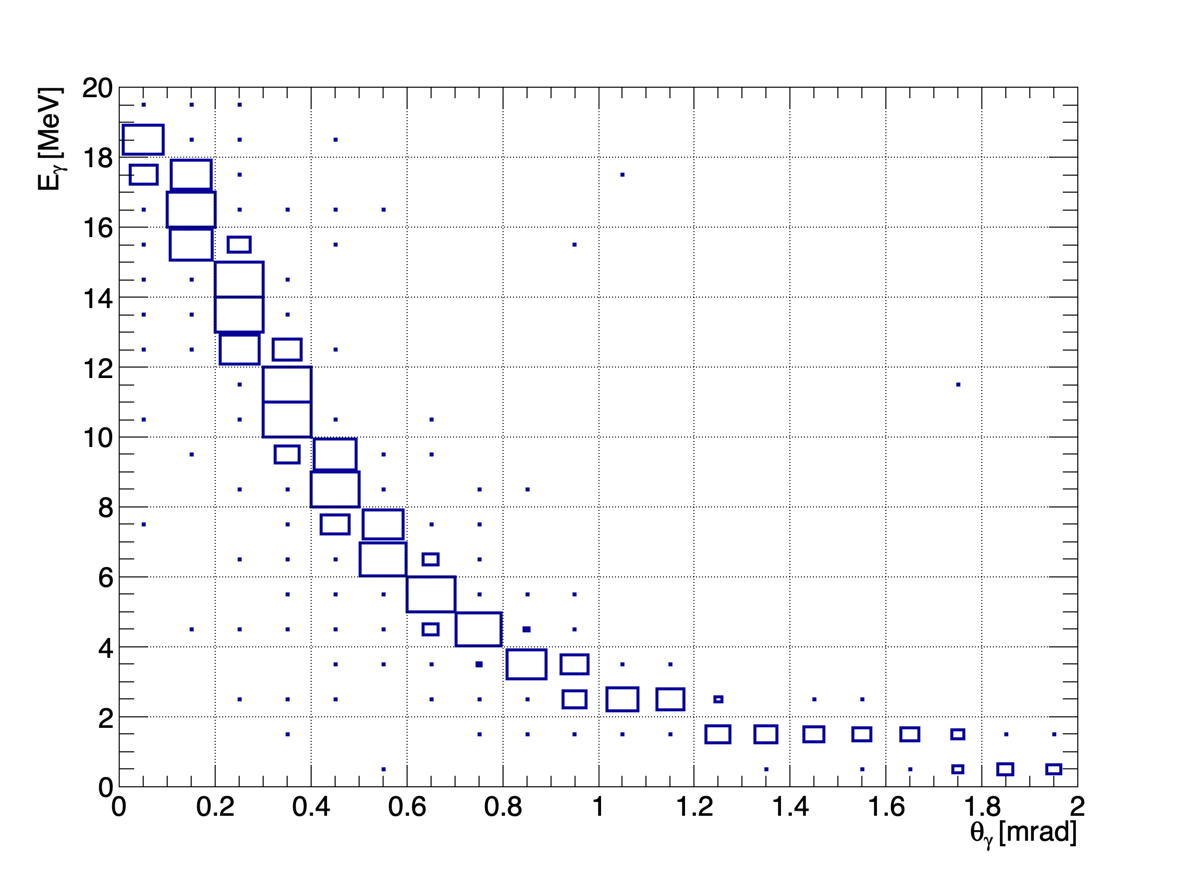}
  \caption{Energy--angle correlation for $\gamma$ emission by ${}^{40}_{20}\mathrm{Ca}^{18+}$ PSI beam at 89.4 TeV.}
  \label{fig:GF_ExT}
\end{figure}

The settings detailed above were chosen to obtain a $\gamma$ beam with a maximum energy of $E_{\gamma}^{max}\,=\,18.7\,$MeV, which covers the photo-fission Giant Dipole Resonance (GDR) of usual actinide targets, like ${}^{238}_{92}\mathrm{U}$ or ${}^{232}_{90}\mathrm{Th}$ \cite{Caldwell:1980}. The energy--angle correlation for the $\gamma$-rays emitted by the GF source with these parameters is shown in Figure~\ref{fig:GF_ExT}.

The dependence shown in Figure \ref{fig:GF_ExT} is centered on a function $E_{\gamma}(\theta_{\gamma})$ given by 
\begin{equation}\centering
E_{\gamma} = \frac{4\gamma_L^2 E_l}{1+\gamma_L^2 \theta_{\gamma}^2}\,,
\label{eq:Energy}
\end{equation}
which is derived from the energy-momentum conservation of the interaction between the ion with the relativistic Lorentz factor $\gamma_L = E/(mc^2)$ and the laser photon with the energy $E_l = \hbar \omega_l$. The spread around this central function is generated by the parameter distributions around their mean values of both the PSI and the laser beams, as listed in Table~\ref{tab:CainPars}.

The maximum $\gamma$-ray energy corresponds to a forward emission angle $\theta_{\gamma} = 0$ and is set by varying the PSI beam energy $E$:
\begin{equation}\centering
E_{\gamma}^{max} = 4\gamma_L^2 E_l = 4\,\frac{E^2 E_l}{m^2c^4}\,.
\label{eq:EnergyMax}
\end{equation}

Together with the possibility to set the minimum energy by collimation, this leads to an important property of a PSI-driven $\gamma$ beam, namely that its energy range is tunable. For example, the GDR range of approximately $10\,$--$\,18\,$MeV is covered by collimation of the $\gamma$ beam to $\theta_{max} \approx 0.4\,$mrad and by tuning the $\gamma_L$ and $E_l$ beam properties to those listed in Table~\ref{tab:CainPars}. LCB-driven gamma beams have the same energy--angle correlation, with the notable difference that a given $E_{\gamma}^{max}$ is reached at much lower electron-beam energy due to the low mass of the electron. Another common feature of these two $\gamma$-beam generation processes is the very narrow angular divergence $\sigma \sim 1/\gamma_L$, which allows for very high fluxes on photo-fission targets. On the other hand, the bremsstrahlung-driven $\gamma$ beams have significantly weaker energy--angle correlations and much broader angular divergences, making them suitable as primary beams for RIBs, only when placed very close to the targets.

Finally, it should be noted that higher energy $\gamma$-rays driven by the PSI beams can be obtained at the LHC. For example, a ${}^{207}_{82}\mathrm{Pb}^{80+}$ beam accelerated to $\gamma_L = 2887$ and a laser beam with $\lambda_l = 104.4\,$nm generate the $\gamma$ beam with $E_{\gamma}^{max} = 396\,$MeV. This wide $E_{\gamma}^{max}$ range, coupled to the tunability of the energy interval, provides a handle on the fission channels activated by the $\gamma$ beam. More specifically, low energy fission in the GDR region is dominated by asymmetric fission. However, as the excitation energy of the fissioning nucleus increases, symmetric fission probability slowly grows and becomes dominant in the delta resonance region around $300\,$MeV.

\section{RIB formation with ion catchers}

The GF gamma beam, tuned to cover the GDR photo-fission energy range, is used to irradiate an actinide target, like ${}^{238}\mathrm{U}$ or ${}^{232}\mathrm{Th}$. In order to generate a significant fission rate and due to the relatively low integrated photo-fission cross section $\sigma_{\gamma f}({}^{238}U) \approx 1\,$b$\,\cdot\,$MeV \cite{Caldwell:1980}, this target needs to be thick (a couple centimeters).
Fission fragments are released from thick targets in ISOL-type facilities by evaporation after heating them to around $2000\degree\,$C. Examples of the ISOL facilities are ISOLDE \cite{Voulot:2008}, which induces fission with a proton beam, and ALTO \cite{Mhamed:2020}, which induces fission with a bremsstrahlung $\gamma$ beam. Although the ISOL method generates high fission rates, it has two shortcomings. The first one is that a large fraction of fission products are refractory elements, with evaporation temperatures in the $3500\,$--$\,4500\degree\,$C range, and are not released from thick targets. The second is that thermal diffusion is a relatively slow process and, in the constant push to populate RIBs with ever more exotic short-lived nuclides, it has become difficult to use.

The solution is to split the thick target in many thin (a few microns) targets.  Fission fragments are then released kinematically, due to their large initial kinetic energy of $50\,$--$\,120\,$MeV. The fragment release is now element-independent and practically instantaneous. Since their release energy is still significant, they need to be slowed down in a gas enclosing the targets which fills the reaction chamber, called 'gas-cell'. After thermalization, the fragments are extracted from the cell by the gas flow through a narrow exit nozzle where a supersonic gas jet is formed. This is the IGISOL production method, first used at JYFL \cite{Aysto:2001} and broadly used at many RIB facilities. The remaining drawbacks of this method is the large extraction time $\sim 100\,$--$\,200\,$ms and small extraction efficiency $<1\%$ of heavy ion transport with gas flow. This is especially true for large gas cells such as the one needed in the case studied here.

The last step in designing an appropriate gas cell is to extract the heavy ions produced in photo-fission with electric fields, rather than with gas flow, drastically improving the transport time and efficiency. This involves a volume DC field that drifts the heavy ions in thermal motion throughout the gas towards the wall with the exit nozzle. A surface RF field on that wall is also needed to catch and transport them towards the nozzle. Therefore, specialized electric devices, called radio frequency carpets (RFCs) \cite{Wada:2003}, are placed on the cell exit wall. They are high density concentric electrodes on printed circuit boards with RF voltages with the frequency $\nu = 2\,$--$\,8\,$MHz adjusted to catch the incoming ions. The transport towards the center is done either with a transverse DC field or with a traveling wave \cite{Bollen:2011}. At the nozzle, the ion extraction is still done by the supersonic gas jet, as with the original IGISOL method.

The ion catchers using electric fields have two potential issues to overcome. The first is the neutralization of the thermalized heavy ions by electron capture. This does not happen with the He gas, which is typically used by the ion catchers, due to its high ionization potential. Molecules of gas impurities can, however, easily release electrons to the heavy ions with which they collide. Therefore, the inert stopping gas has to be kept at a very high-purity level of around $1\,$ppb. The second issue is the ionization of the He gas by the energetic heavy ions during their slowdown, leading to the formation of a He$_{3}^+$ ionic cloud. When it becomes strong enough, this space-charge plasma effect can completely overcome the electric fields created by the electrodes and stop the heavy ion extraction. The space charge effects for the extraction of RIBs in the particular case discussed in this article will be analyzed in the next section.

The particular ion catcher technology proposed for the RIBs production at the GF facility is the high areal density with the orthogonal extraction cryogenic stopping cell (HADO-CSC) introduced in Reference \cite{Dickel:2016}. This device is developed in collaboration by GSI (Germany), the Giessen University (Germany), ELI-NP (Romania), Soreq (Israel) and JYFL (Finland) for the future RIB facilities at FAIR/LEB, ELISOL (ELI-NP) and SARAF (Soreq). The HADO-CSC introduces several technological improvements over the state-of-the-art ion catchers. The gas purity is kept at about $1\,$ppb by continuously recirculating and purifying the He gas with a system of active and passive purifiers and also by cooling the inner chambers to the temperature around $75\,$K, which freezes most of the impurities on the walls. The heavy-ion drift orthogonal to the primary beam, as opposed to the standard longitudinal drift, allows for much higher DC fields without increasing the voltages applied on the electrodes. Due to the high fields, the HADO-CSC can operate at space charge rates with five orders of magnitude higher. There are several other important improvements of this design described in detail in Reference \cite{Dickel:2016}.

\section{Simulation of production rates}

\begin{figure}[h]\centering
  \includegraphics[width=0.52\linewidth]{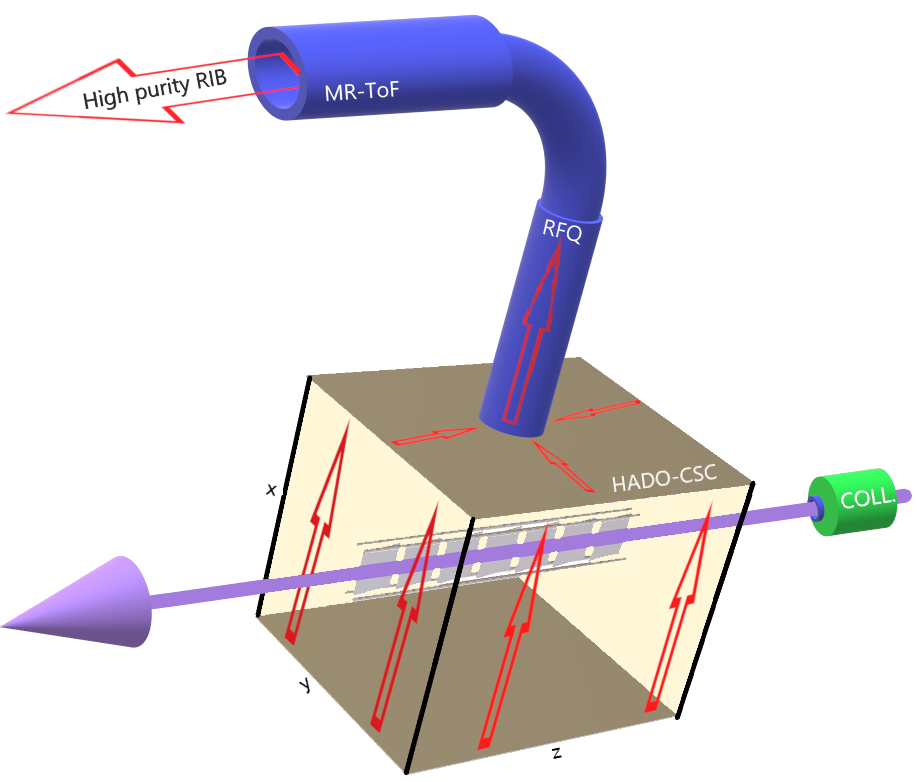}
  \caption{(colors online) ELISOL generic setup scheme depicting the $\gamma$ beam (purple) hitting many thin targets (light gray) placed inside the He filled HADO-CSC. The produced fission fragments are extracted using electric fields (red arrows), formed into a beam by a radio-frequency quadrupole (RFQ) and finally obtain a high-purity RIB using a multi-reflection time of flight (MR-ToF) mass spectrometer.}
  \label{fig:ELISOL}
\end{figure}

For estimating the production rates of an ELISOL-type \cite{Constantin:2017} RIB facility at the GF, we have performed {\sf GEANT4}~\cite{Agostinelli:2002hh} simulations using the beam parameters (the energy--angle correlation, the energy distribution and intensity) and the relevant parts of the ELISOL setup, shown in Figure~\ref{fig:ELISOL}, consisting of a HADO-CSC with dimensions of $25\,$cm$\,\times\, 25\,$cm in orthogonal plane and about $2\,$m in lengths filled with pure He at $300\,$mbar and $75\,$K hosting $54$ UO$_2$ $3\,\mu$m thin circular targets inclined at $10 \degree$ with respect to the beam axis, parameters that were found optimal for the photon-induced RIB creation in the previous study \cite{Nichita:2020}.

\begin{figure}[h]\centering
  \includegraphics[width=0.48\linewidth]{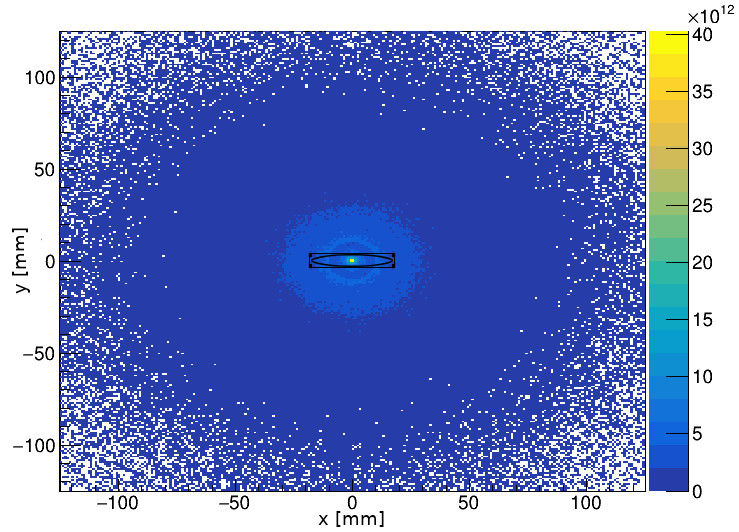}
  \caption{The beam spacial profile hitting the targets (black ellipse) enclosed in their frames (black rectangle) at $100$ meters away from the GF interaction point. The color scale shows the flux [$\gamma$/s/mm$^2$] on the orthogonal plane ($\textit x,y$).
  }
  \label{fig:BeamProfile}
\end{figure}

The proposed distance of the target system with respect to the GF interaction point is $100\,$m with the corresponding beam profile shown in Figure~\ref{fig:BeamProfile}. The beam is collimated such that it completely covers the first target of the setup, thus eliminating the unnecessary $\gamma$-rays that produce just background.

The targets projection in the plane orthogonal to the beam is depicted by the black ellipse in Figure~\ref{fig:BeamProfile}. After collimation, the beam fraction hitting the target system at that particular distance is about $3\%$ of total intensity.

\begin{figure}[h]\centering
  \includegraphics[width=0.98\linewidth]{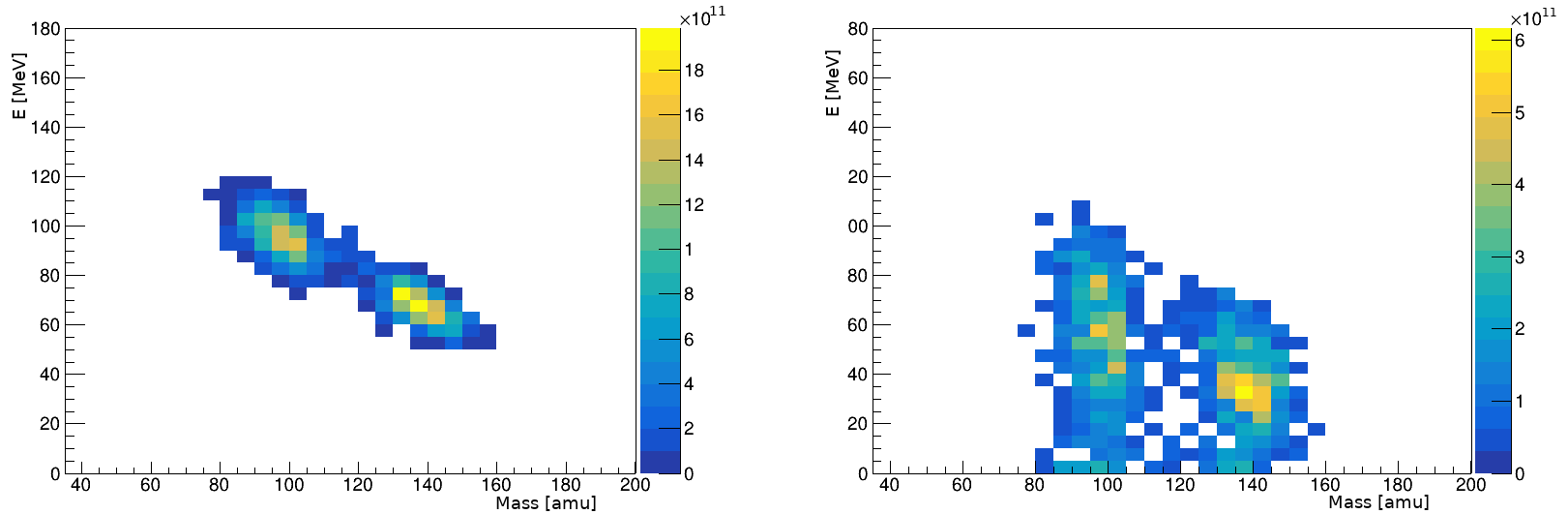}
  \caption{(colors online) Ions energy at the production point (left) and when released from the targets into the gas (right). The color scale shows the intensity [ions/s] for the produced and release ions respectively.}
  \label{fig:IonsEnergy}
\end{figure}

Using the GIF model \cite{Mei:2017} for photon-induced fission and {\sf GEANT4} framework, we estimate the yield, mass and charge distribution and the energy of the produced ions. We track the ions escaping the targets and are released in the gas and compute the energy deposited in He, while they are stopped, using ATIMA model \cite{Weick:2002} inside {\sf GEANT4} framework. In Figure~\ref{fig:IonsEnergy} are shown the energy and mass distributions of the produced and released ions, respectively.     

The {\sf GEANT4}-simulated yield of the ions released in the gas is $10^{12}$ ions/second which is orders of magnitude higher than in the current RIB facilities, but this high rate produces significant space charge inside the cell. While ions are stopped, they deposit their energy in the gas mainly by ionization, creating the He$^+$ ions that arrange themselves into He$_3^+$ trimers. The fission fragments have a mean kinetic energy of $39.8\,$MeV when they escape the targets, and thus each stopped ion is producing about $10^6$ He$^+$-e$^-$ pairs, taking $41\,$eV \cite{Constantin:2017} as the He ionization potential. 

\begin{figure}[h]\centering
  \includegraphics[width=0.65\linewidth]{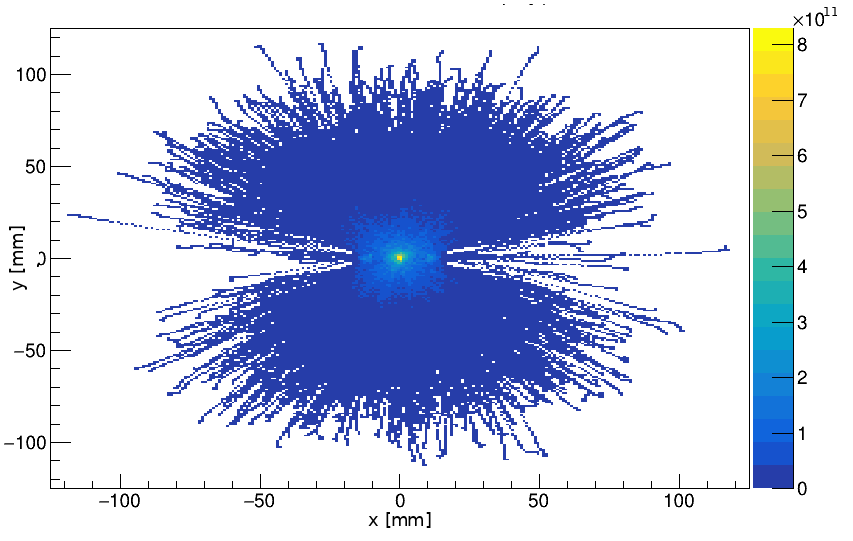}
  \caption{He$_3^+$ ions/second/mm$^2$ flux spacial distribution, in the orthogonal plane, generated due to fission fragments energy deposition in the gas.
  }
  \label{fig:IonStopping}
\end{figure} 

\begin{figure}[h]\centering
  \includegraphics[width=0.60\linewidth]{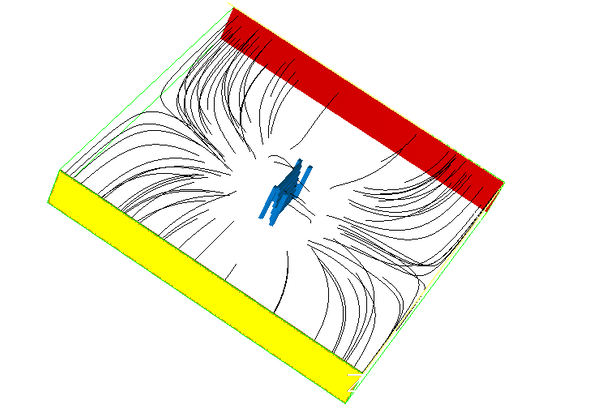}
  \caption{(colors online) The {\sf SIMION} geometry consisting of the full orthogonal plane of the CSC, thick enough to accommodate one UO$_2$ circular target ($D=30\,$mm) embedded in a AlMg$_3$, $35\,$mm$\,\times\,35\,$mm square frame inclined at $10\degree$ with respect to the beam axis and secured with $4$ metal rods ($D=3\,$mm) in the corners. The external field is created between the extraction electrode (red), which is put on a negative potential to attract the positive ions, and the grounded electrode (yellow). The target system (blue) is put at half of the potential of the extraction electrode in order to obtain a smooth field and suppress the very near He$_3^+$ ions.}
  \label{fig:SimionCSCslice}
\end{figure}

The total charge of the He$_3^+$ ions accumulated for $1$ second at the full beam is then about $1\,$mC. We estimate the ions extraction efficiency using {\sf SIMION}~\cite{SIMION} simulations of a slice of the HADO-CSC, a $25\,$cm $\,\times\, 25\,$cm full orthogonal plane, $3.3\,$cm thick, in which we have added one circular target with the corresponding frame (Figure \ref{fig:SimionCSCslice}). The results are relevant for the whole CSC which can be seen as being made up of $54$ identical parts as the one simulated.

\begin{figure}[!h]\centering
  \includegraphics[width=0.57\linewidth]{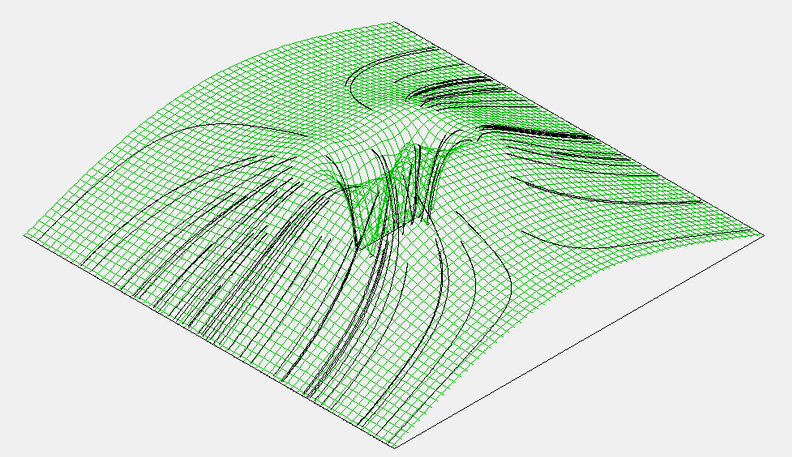}
  \caption{The static simulation with a small number of ions showing the extraction behavior due to the high space charge density in the center of the CSC on top of the total electric field created (external plus space charge).}
  \label{fig:StaticSim}
\end{figure}

We proceed for static simulations using variations of the extraction electric field from $100\,$V/cm up to $500\,$V/cm, which is approximately a half of the Paschen's Law limit, and superimposing the full space charge imported from {\sf GEANT4}. As expected, the formed He$_{3}^{+}$ cloud density shown in Figure~\ref{fig:IonStopping} is focused in the center of the cell in the vicinity of the target system, where the ions have most of energy and a higher spacial density. Due to this high positive charge created in the center, the ions are pushed and accelerated towards the cell walls with high velocity, with behavour shown in Figure \ref{fig:StaticSim}. 

\begin{figure}[!h]\centering
  \includegraphics[width=0.7\linewidth]{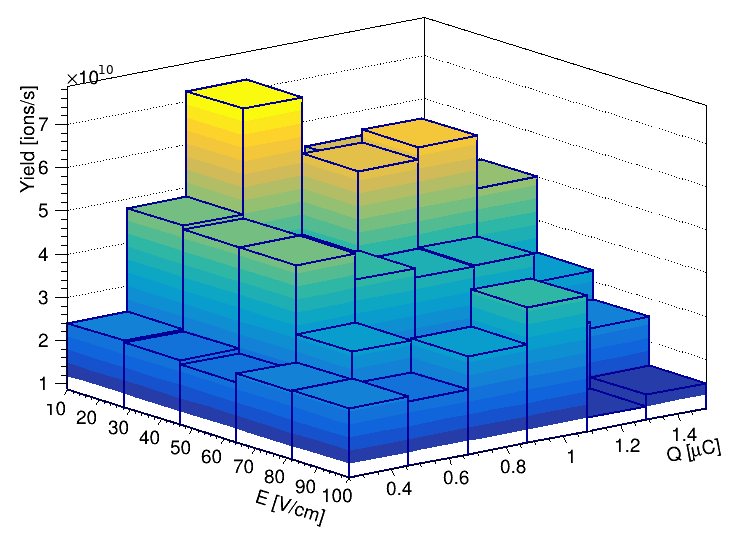}
  \caption{(colors online) The total extracted ion yield for variations of the external electric field (E) and the total space charge in the cell (Q). }
  \label{fig:IonYield}
\end{figure}

To prevent the ions from penetrating the cell wall, and thus loose them, strong but short repulsive electric fields are used, generated by radio-frequency circuits, called the RF carpets \cite{Wada:2003}. The current state-of-the-art RFCs can catch ions that have a perpendicular velocity lower than $50\,$m/s \cite{Rotaru:2019}. In our static simulation, the high space charge, which is located mostly in the center, acts as an electrode with high potential that pushes the ions towards the walls. An important fraction of the ions, about $19\%$, is estimated to hit the extraction electrode, but due to the high velocity (more than $5\,$km/s) they cannot be stopped and cached by the RF field, and thus escape by penetrating the RF carpet that leads to almost no extraction.
The need to lower the space charge while keeping a maximal ion yield, led to the idea of chopping the beam in such a way that the space charge is accumulated for much less time and then evacuated before the next bunch. The {\sf SIMION} estimation for the extraction time is less than $1\,$ms, thus the proposed solution is to chop the beam in $1\,$ms on and $1\,$ms off bunches, resulting in a space charge reduction by $3$ orders of magnitude, while the ion yield is lower just by a factor of $2$ to the rate of $0.5\times 10^{12}$ ions/second. This important reduction of the space charge is due to the fact that the ionization is accumulated for just $1\,$ms versus $1\,$s, and then in the next $1\,$ms (when the beam is off) there is time that the charge is fully evacuated and the cell is ready for the next bunch. The ion yield is cut just by half because the ion production takes place with the full beam intensity for periods equal to the paused period.
With a total space charge of $10\,\mu$C, corresponding to the described chopped beam, we have performed full static and also dynamic, particle-in-cell (PIC) {\sf SIMION} simulations with the up and down space charge and external extraction field variations in order to find the maximum yield for the extracted ions (Figure \ref{fig:IonYield}).

\begin{figure}[!htbp]\centering
  \includegraphics[width=0.50\linewidth]{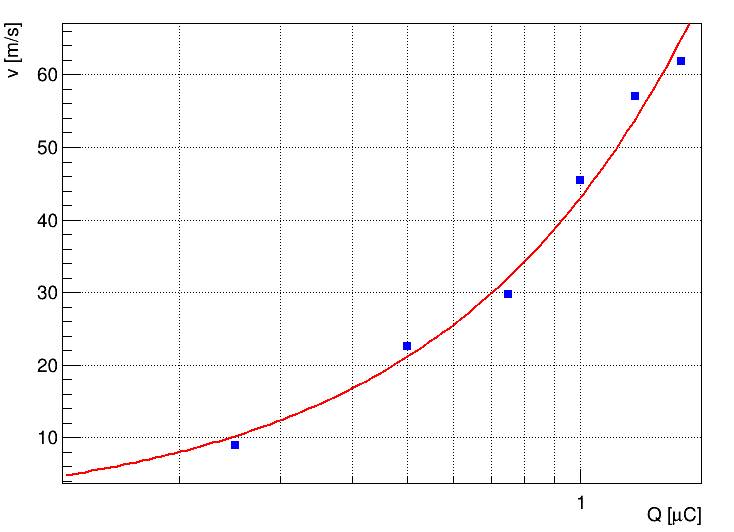}
  \caption{Mean velocity of the ions versus the total space charge in the cell (blue squares) and fit function (red line)}
  \label{fig:IonVelocity}
\end{figure}

In this range of the space charge, the mean velocity of the ions is near the RF carpet limit of $50\,$m/s and for the optimum case the velocity is safely below this limit, as shown in Figure~\ref{fig:IonVelocity}. The external field plays the role of guiding the ions towards extraction and usually a stronger field is preferred for improving the extraction time and efficiency. In the presented case, the ions are strongly pushed to the extraction electrode by the high centered density of space charge, thus a strong external field would increase the velocity of the ions above the $50\,$m/s limit when hitting the extraction RF carped. The best extraction efficiency of about $8\%$ has been obtained using a relatively low, $10\,$V/cm, external field combined with the $0.8\,\mu$C total space charge accumulated in the cell, corresponding to a beam bunch time profile of $0.8\,$ms on and $1\,$ms off. 

An even lower external field could not be used due to the electron--ion recombination effect. 

\begin{figure}[!htbp]\centering
  \includegraphics[width=0.57\linewidth]{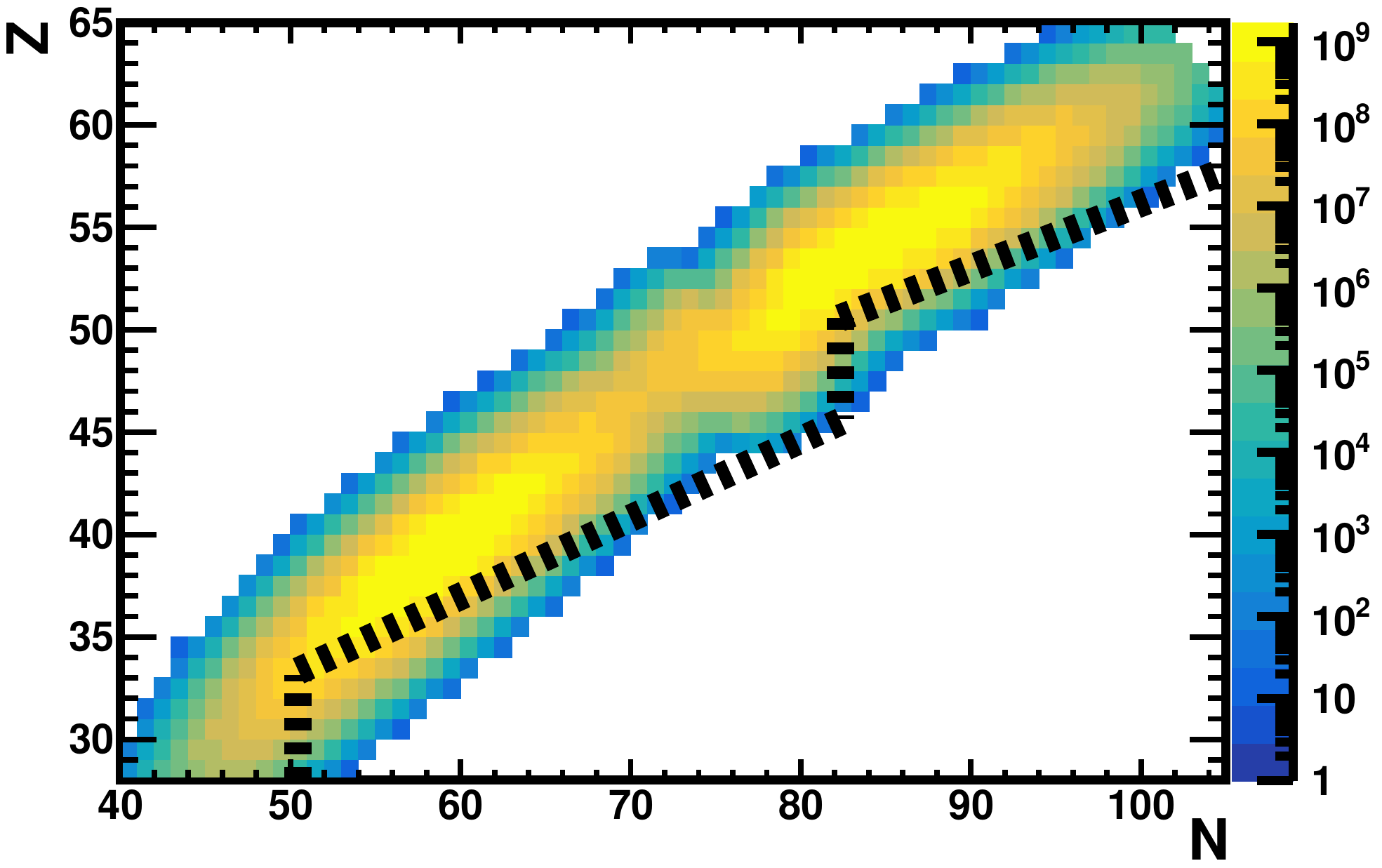}
  \caption{(colors online) The extracted ions-yield (side color scale in ions per second) as a function of proton number (Z) and neutron number (N) isotope distribution chart and the $r$-process line (dotted black line).}
  \label{fig:YZN}
\end{figure}

The high ion yield of about $7.2\times10^{10}$ ions/second, extracted from the proposed setup, permits the measurement and analysis of rare exotic nuclei, which are produced with a very low cross section, and covers both the $N=50$ and $N=82$ $r$-process waiting points. Figure~\ref{fig:YZN} is generated using a fit of the distribution of the ions being released from the targets resulting from the {\sf GEANT4} simulations, normalized to the extracted ions/s yield presented as the optimum case in this work with the assumption that the extraction process is isotope-independent.
The plot is stopped at $100$ ions/s because statistical fluctuations become
large below that point.

\section{Conclusions}
The proposed ELISOL type setup at the GF produces ions yields of the order $10^{12}$ ions/second via photo-fission process of the high flux $\gamma$-rays impinging on multiple thin actinide targets. This production rate correlated with the proposed solutions to lower the space charge effect are resulting in extracted ion yields of about $0.72\times10^{11}$ ions/second which is orders of magnitude higher than the current RIB facilities. Although the assumption for isotope-independent extraction is not quite precise, the yield distribution which is presented in Figure \ref{fig:YZN} demonstrates the big potential of the technique for studies of neutron-rich nuclei, lying far away from the valley of $\beta$-stability. These results clearly demonstrates the opportunity to produce high-intensity RIBs at the GF and make available, for laboratory studies, RIBs of most exotic isotopes and refractory elements which are kinetically released into gas from the thin targets. Another key advantage of the technique is that due to the ultra fast electric field extraction, estimated to be less than a few milliseconds, one can measure the very short lived nuclei with higher precision than achieved before.



\medskip

\medskip
\textbf{Acknowledgements} \par 
The authors would like to thank T. Dickel, W. Pla$\ss$ and C. Scheidenberger for the excellent collaboration in the HADO-CSC project. This work is supported by Extreme Light Infrastructure-Nuclear Physics (ELI-NP) Phase II, a project co-financed by the Romanian Government and by the European Regional Development Fund the Competitiveness Operational Programme (1/07.07.2016, COP, ID 1334).

\medskip

%

\bibliographystyle{MSP}
\bibliography{RIB}

\end{document}